# Anisotropic Electron-Photon-Phonon Coupling in Layered MoS$_2$

Deepu Kumar[1#], Birender Singh[1], Rahul Kumar[2], Mahesh Kumar[2] and Pradeep Kumar[1*]

[1]School of Basic Sciences, Indian Institute of Technology Mandi, 175005, India

[2]Department of Electrical Engineering, Indian Institute of Technology Jodhpur, Jodhpur-342037, India

**Abstract:**

Transition metal dichalcogenide, especially MoS$_2$ has attracted lot of attention recently owing to its tunable visible range band gap and anisotropic electronic and transport properties. Here, we report a comprehensive inelastic light scattering measurements on CVD grown (horizontally and vertically aligned flakes) as well as single crystal flakes of MoS$_2$, probing the anisotropic optical response via studying the polarization dependence intensity of the Raman active phonon modes as a function of different incident photon energy and flake thickness. Our polarization dependent Raman studies intriguingly revealed strong anisotropic behavior reflected in the anomalous renormalization of the modes intensity as a function of flake thickness, phonons and photon energy. Our observations reflects the strong anisotropic light-matter interaction in this high crystalline symmetric layered MoS$_2$ system especially for the in-plane vibrations, which is crucial for understanding as well application of these materials for future application such as optoelectronic applications.

[#]E-mail: deepu7727@gmail.com

* E-mail: pkumar@iitmandi.ac.in

# 1. Introduction

The journey of two dimensional (2D) nano materials begins with the discovery of Graphene [1], the first 2D material defying the Mermin-Wagner theorem. Since the discovery in 2004, these material have been intensively probed both experimentally as well as theoretically owing to their extraordinary properties and rich physics [2]. Recently, it has been joined by the discovery of large class of other 2D materials namely 2D transition metal dichalcogenides (TMDCs). In the bulk form TMDC have been probed for decades [3-4] but the current interest in these materials is renewed because of their atomically thin 2D form. TMDCs have the properties which are complementary to those of graphene such as graphene is a zero gap semiconductor hence it can't be switched off and as a result not good for semiconductor industry. On the other hand TMDCs provide a tunable band gap thus overcoming the bottleneck faced by graphene and are promising materials for next generation electronic and optoelectronic applications. In these systems, one may tune their properties as a function of layer thickness providing great leverage against the conventional 3D bulk systems. Additionally, numerous properties of TMDCs such as electrical mobility, photoluminescence, photo-responsivity and thermoelectric figure of merit show strong dependence on the in-plane crystalline orientation [5-8] as well as layer thickness. Interestingly, all these properties may be intriguingly linked with the underlying behavior of the electrons, phonons and quasi-particles, which may be investigated using light-matter interaction such as Raman spectroscopy.

$MoS_2$ is one of first 2D TMDC system being explored as an alternative to replace silicon for semiconductors industry owing to the tunable band gap, from ~1.2 eV in bulk to 1.9 eV for monolayer [9]. In addition, to the finite spin-orbit coupling, broken inversion symmetry in $MoS_2$ provide optical access to the valley degrees of freedom suggesting its potential in future spintronic and valleytronic devices. Light-matter interaction in these 2D materials provide a deep insight into the underlying interaction mechanism as photon may easily interact with electrons and phonons, which control their electrical and thermal properties. The underlying mechanism responsible for their stark difference in physical properties as one moves from bulk to monolayer may be liked with the behavior of quasi-particles and coupling between them such as anisotropic coupling between electrons and phonons/photons. This anisotropy may be gauged via an indirect probe by an external perturbation such as probing by light, in particular Raman scattering. Raman scattering has proved to be a very powerful technique to understand intricate coupling between different quasi-particles in nano systems as well as bulk via coupling of electrons and phonons with the incoming and outgoing photons [10-15].

Phonon modes in TMDC are sensitive to the thickness of these materials. In case of MoS$_2$ as one moves from bulk to monolayer then out of plane phonon mode $(A_{1g})$ involving the vibration of S atom is red-shifted; on the other hand $E^1_{2g}$ mode involving the in-plane displacement of Mo and S atoms shows blue-shift attributed to the dialectic screening of long-range coulombic interaction [15]. Interestingly in 2D materials, phonon modes may be tuned by varying the incident photon energies if it falls within the resonance range of optical transition. In fact close to the resonance condition, one may get additional modes including the forbidden one owing to the broken symmetry and/or resonance effect which eases the momentum conservation rule. In case of MoS$_2$ excitation with 633 nm (1.96 eV) laser, which is close to *A* exciton energy, gives rise to many multiphonon modes. These multimode provide very rich information about phonons in the entire Brillouin zone (BZ) as opposed to the conventional Raman process where observed phonon modes are confined only at the Γ point in the BZ. A detailed study focusing on the polarization dependence of phonon modes as a function of incident photon energy and layer thickness was lacking for high symmetric MoS$_2$. Focusing on the polarization dependence of the phonon modes, here, we have undertaken such a study and report a comprehensive study in vertically and horizontally CVD grown MoS$_2$ as well as mechanically exfoliated MoS$_2$ from bulk single crystal as a function of flake thickness and different incident photon energies. We show how the electron-photon and electron-phonon interaction in MoS$_2$ are related to its thickness reflecting via the different polarization dependence behavior of phonon modes a function of layer thickness as well as incident photon and phonon energy.

## 2. Experimental Techniques

Vertically and horizontally aligned MoS$_2$ was synthesized using CVD as described in ref. 16. Single crystal flakes were obtained using scotch tape method from the bulk single crystal of MoS$_2$ (2D semiconductors, USA) on 300nm SiO$_2$/Si substrate. Polarization dependence Raman spectroscopy measurements were done by fixing the direction of polarization of the incident beam as well as rotation of the sample, but have made use of the scattered light polarization direction using analyzer (see Fig. 1 for the exp. geometry); and all the measurements were done in backscattering geometry using 532 nm and 633 nm laser. The spectra were collected using a 100X long working distance objective and Peltier cooled Charge Couple Device (CCD) detector. The laser power was kept very low (~ 500 $\mu$W) to avoid any heating.

## 3. Result and Discussions

TMDCs have chemical formula MX$_2$, where two nearest layers of X (chalcogen atom) are separated by a layer of M (transition metal), actually the monolayer of MX$_2$ system is composed of a trilayer i.e. X-M-X. For bulk MoS$_2$ (point group $D_{6h}^4$), Mo is sandwiched between two S atom layer forming hexagonal plane and Mo is also connected with S via a trigonal prismatic coordination in 2H phase, where H stands for the hexagonal symmetry and prefix 2 for the number of layers in each stacking order i.e. after two layer pattern is repeated [3]. As one move from bulk to finite number of layer then symmetry reduces owing to lack of translational symmetry in Z direction; 24 symmetry operation in bulk reduced to 12 in few layer and as a result space group is different for the few layered TMDCs systems (see table-S1). Monolayer/odd number of layer of MoS$_2$ belongs to $D_{3h}^1$ point group $(P\bar{6}m2, \#187)$, on the other hand bilayer/even number of layer belongs to $D_{3d}^3 (P\bar{3}m1, \#164)$ [17-19]. Details of the character tables, Raman tensor of the phonon modes at the $\Gamma$ point and irreducible representation are given in detail in Table-S1 for all the three groups discussed above [20-22].

Figure 2 shows the Raman spectra of CVD grown horizontally aligned MoS$_2$ for two incident photon energies (i.e. 532 nm and 633 nm) with increasing thickness of the MoS$_2$ at four different flakes. Two first-order Raman active phonons modes labeled as $A_{1g}$ and $E_{2g}^1$ are observed around 380 and 405 cm$^{-1}$; respectively for all the flakes. Additionally, we observed few other modes for 633 nm incident photon energy attributed to the higher-order overtone modes owing to the resonance effect. From the frequency difference ($\Delta\omega = \omega_{A_{1g}} - \omega_{E_{2g}^1}$) of mode $A_{1g}$ and $E_{2g}^1$, one may estimate the number of layers in these flakes. For monolayer, the frequency difference ($\Delta\omega$) is ~18 cm$^{-1}$ for mechanically exfoliated one and ~ 20 cm$^{-1}$ for CVD grown sample, and difference increases by ~ 2-3 cm$^{-1}$ for the case of bilayer and with further addition of layers it increase by ~ 1 cm$^{-1}$ per layer till six-seven layered [23-24].

To understand the anisotropic coupling between electrons, photons and phonons, a compressive polarized Raman scattering measurements were carried out on CVD grown horizontally aligned

MoS$_2$ flakes. Figure 3 shows the polarization dependence of $A_{1g}$ and $E^1_{2g}$ modes for both 532 and 633 nm laser as a function of increasing thickness of horizontally aligned flakes of MoS$_2$. For flake 1 (532 nm laser), the intensity of $E^1_{2g}$ mode remains invariant with respect to the rotation of analyzer, however intensity of $A_{1g}$ mode shows two-fold symmetric nature i.e. it has maximum intensity at $0^0$ and $180^0$. One may understand the observed variation in the intensity as a function of the analyzer angle within the semi-classical approach. As incident polarized light lies in the XY plane, polarized vector ($\hat{e}_i$) of the incident beam may be decomposed as $(\cos\alpha, \sin\alpha, 0)$, where $\alpha$ is an arbitrary angle from x-axis (see Fig.1d ). We have fixed the direction of $\hat{e}_i$ and rotated the analyzer at an interval of $10^0$, polarization vector ($\hat{e}_s$) for the scattered light also lies within the XY plane with decomposition as $[\cos(\theta+\alpha), \sin(\theta+\alpha), 0)]$ where $\theta = 0^0$ to $360^0$. Within the semi-classical approach, Raman scattering intensity is given as $I_{int} = |\hat{e}^t_s.R.\hat{e}_i|^2$, where t denotes the transpose of $\hat{e}_s$ and R represents the Raman tensor [25]. Using the above expression intensity of the $A_{1g}$ and $E^1_{2g}$ mode for our experimental setup is given as $I_{A_{1g}} = a^2 \cos^2\theta$, $I_{E^1_{2g}} = d^2(\cos^2\theta + \sin^2\theta)$; respectively. $I_{A_{1g}}$ decreases to zero when $\hat{e}_s$ and $\hat{e}_i$ are perpendicular to each other and is maximum for $\theta = 0$ i.e. when $\hat{e}_s$ and $\hat{e}_i$ are parallel; on the other hand $I_{E^1_{2g}}$ remains constant. Solid lines in Fig. 3 (Flake 1, 532 nm) are the fitted curve using the above functions and the fitting is in very good agreement with the theoretical prediction.

As we changed the energy of incident photon (to 633 nm) and considering the data on the same flake and same spot (see Fig. 3, 633nm, flake 1), remarkably intensity of the $E^1_{2g}$ mode is no longer invariant under rotation but shows that the intensity is maximum around $0^0$ and $180^0$ similar to that of $I_{A_{1g}}$. As we moved to the next flakes (flakes # 2,3 and 4) for 532 and 633 nm laser, intensity pattern of the $A_{1g}$ mode remains constant. Interestingly for 532 nm laser, isotropic nature of $I_{E^1_{2g}}$ changes slightly with increasing thickness i.e. the constant intensity is showing signature of minima around $90^0$ and $270^0$; on the other hand as we changed the laser to 633 nm the intensity pattern for $E^1_{2g}$ mode shows remarkable changes and becomes two-fold symmetric similar to that of $A_{1g}$ mode. Solid lines for $A_{1g}$ (both 532 and 633 nm) and $E^1_{2g}$ mode (for 633 nm) are fitted with the functions $a^2\cos^2\theta$ and the fitting is very good. For $E^1_{2g}$ mode for 532 nm, where the intensity pattern is deviating from constant value and forming a semi-lobes kind of structure, we have fitted using the combined functions i.e. $(d^2+a^2)\cos^2\theta+d^2\sin^2\theta$, overall fitting is modest. Figure 4 shows the intensity ratio ($I_{E^1_{2g}}/I_{A_{1g}}$), and the intensity ration diverges as one moves close to $90^0$ (see Fig.4a) on the expected lines within the theoretical prediction, on the other hand intensity ratio in Fig. 4b shows constant behavior with respect to the rotation of analyzer completely opposed to that observed in Fig.4a.

Figure 5 shows the polarization dependence data for $A_{1g}$ and $E^1_{2g}$ mode for both 532 and 633 nm laser as a function of increasing thickness, for four different flakes of vertically aligned MoS$_2$. The intensity of the out of plane vibrational mode ($A_{1g}$ mode) shows consistent behavior irrespective of the incident photon energy and layer thickness, except some minor changes in the lobes sizes; for example lob size in the 2$^{nd}$-3$^{rd}$ quadrant is more expanded for flake 1 and 4 (for 532 nm). On the other hand intensity of the in-plane vibrational mode ($E^1_{2g}$) changes drastically as we vary the incident photon energy as well as layer thickness, similar to what we observed for the horizontally aligned MoS$_2$ discussed above. The observed asymmetry reflected in the in-plane vibrational mode ($E_{2g}^1$) clearly suggest strong anisotropic nature of the phonon modes in MoS$_2$, and weaker polarization dependence of out of plane vibrational ($A_{1g}$) as compared to in-plane vibrational mode ($E_{2g}^1$), suggesting that out of plane vibrations couple less anisotropically with the electronic states as compared to the in-plane vibrations. Intensity ratio (see Fig. S1 for the Raman spectra as a function of flake thickness and Fig. S2 for the intensity ratio) also shows similar changes as observed for the case of horizontally aligned MoS$_2$ discussed above. To probe this anisotropic coupling further, we also did similar measurements on single crystal flakes of MoS$_2$ at five different flakes starting from monolayer of MoS$_2$ (see Fig. 6). Interestingly single crystals flakes also show similar polarization-dependent behavior as that observed for the horizontally and vertically aligned flakes i.e. intensity of the $A_{1g}$ mode shows two fold symmetric nature irrespective of the incident photon energy and layer thickness; while intensity of $E_{2g}^1$ changes as a function of layer thickness as well as incident photon energy. Intensity ratio (i.e. $I_{E_{2g}^1}/I_{A_{1g}}$) also shows similar behavior as for horizontally and vertically aligned MoS$_2$ (see Fig. S3 for the Raman spectra as a function of flake thickness and Fig. S4 for the intensity ratio). The results shown in Fig. 3, 5 and 6 clearly suggests strong anisotropic optical response in this high crystalline symmetric layered MoS$_2$ and its intricate dependence on the incident photon energy, phonon energy as well as flake thickness.

From our observation on these three different forms of MoS$_2$ as a function of layer thickness as well as incident photon energies, it evidently reflects the anisotropic nature of the phonons. It is also clear that the semi-classical approach can't capture the complete picture with respect to the observed anisotropies especially for the case of in-plane vibrations. For example, with in this semi-classical approach a phonon mode with the same symmetry would have the same polarization dependence irrespective of the incident photon energy; what we observed for $E_{2g}^1$ mode (see Fig. 3, 5 and 6) is completely different from this prediction. This failure of semi-classical approach to explain the polarization dependence as a function of layer thickness and incident photon energy ($E_L$) may be understood from the fact that the intensity of a Raman active mode within this approach is independent of the $E_L$ as well as the thickness of the materials. This independence of the intensity from $E_L$ and thickness comes from the fact that the optical dipole selection rule for the absorption and emission of a photon is not included in this semi-classical theory. All it depends upon is the direction of polarization of the incident, scattered photons via $\hat{e}_s$, $\hat{e}_i$ and Raman tensor (R), which is related to the rate of change of polarization for a given mode i.e. it depends on the symmetry of a vibration. To qualitatively understand these anisotropies in the phonon modes as a function of layer thickness and $E_L$; quantum mechanical picture needs to be invoked, which depends on the interaction between electron-photon-phonon. In a simplistic picture, stokes Raman scattering process may be understood via three steps as follow (also diagrammatically depicted in Fig. 1a, b):

(i) When a laser impinges on the system, the electronic system absorb a photon and electron-hole pair is created resulted into finite electron-photon interaction ($H_{op}$)

(ii) Electron goes to the excited state and it creates a phonon (stokes process) into the system and results into electron-phonon interaction ($H_{el-ph}$).

(iii) Finally, the electron comes back to the initial state and combine with hole and emit the photon (electron- photon interaction; $H_{op}$).

From the above picture describing the Raman scattering process one may easily gauge that Raman process can play a pivotal role in deciphering the intricate coupling between photon, phonons and electrons. The first-order Raman scattering process involves three steps, as descried above, and quantum mechanical expression for the Raman intensity, which uses the dipole selection rule for the optical transition, is given as [25-26]

$$Int. = \left| \sum_{g,i,i'} \frac{\langle g|H_{op}|i'\rangle \langle i'|H_{el-ph}|i\rangle \langle i|H_{op}|g\rangle}{(E_L - \Delta E_{ig})(E_L - \hbar\omega_{ph} - \Delta E_{i'g})} \right|^2 \quad \text{------- (1)}$$

where $\Delta E_{ig} = (E_i - E_g + i\gamma)$ and $\Delta E_{i'g} = (E_{i'} - E_g + i\gamma)$

Where $|g\rangle$ is the ground state, $|i\rangle/|i'\rangle$ are the intermediate states (eigen or non-eigen one), $H_{op}$ and $H_{el-ph}$ are the electron-photon and electron-phonon interaction; respectively. $E_L$ is the incident photon energy and $E_g$, $E_i$ and $E_{i'}$ are the energy of the corresponding electronic states, and $\gamma$ is the broadening factor which is introduced phenomenologically for the finite life time of the absorption/emission. Matrix element $\langle i|H_{op}|g\rangle$ and $\langle g|H_{op}|i'\rangle$ represent the process of optical absorption and emission; respectively and $\langle i'|H_{el-ph}|i\rangle$ is the electron phonon matrix element. Treating the electron-photon interaction within the dipole approximation i.e. by neglecting higher order multipole or non-linear effects, the electron-photon matrix element $\langle i|H_{op}|g\rangle$ is given as $\langle i|H_{op}|g\rangle \propto \hat{e}_i.\vec{D}_{ig}$, where $\hat{e}_i$ is the polarization vector of the incident light and dipole vector $\vec{D}_{ig}$ is $\langle i|\nabla|g\rangle$ [27-28]. From this expression, it is clear that to have a non-zero contribution of electron-photon matrix element $\vec{D}_{ig}$ should have a finite component along the direction of light polarization $\hat{e}_i$. This may also explain, which two states ($|g\rangle, |i\rangle$) from the energy bands be involved in this transition of an electron for a given $E_L$ and direction of $\hat{e}_i$, which explains the different possible values of the matrix element $\langle i | H_{op} | g \rangle$ for varying incident light polarization direction and incident photon energies. As there are three matrix component involved (see eq$^n$. 1 above), to get finite intensity for a mode each component should be non-zero and be allowed by the symmetry. For e.g. to obtain non-vanishing electron-photon matrix elements for the optical transition from ground state to the intermediate state $|i\rangle$, it should satisfy $\Gamma_g \otimes \Gamma_D \subset \Gamma_i$, where $\Gamma_D$, $\Gamma_g$ and $\Gamma_i$ are the irreducible representations of $\vec{D}_\updownarrow$ (i.e. parallel component of $\vec{D}_{ig}$ with respect to $\hat{e}_i$), $|g\rangle$ and $|i\rangle$, respectively. Similarly $\langle g | H_{op} | i' \rangle \propto \hat{e}_s . \vec{D}_{gi'}$, where $\hat{e}_s$ is the polarization vector of the scattered light and $\vec{D}_{gi'} = \langle g | \nabla | i' \rangle$. Comparing quantum expression (eq$^n$.1) with the semi-classical, one may write the XY component of Raman tensor, say for $E^1_{2g}$ mode as $d \propto [\langle g | \frac{\partial}{\partial y} | i' \rangle \langle i' | H_{el-ph} | i \rangle \langle i | \frac{\partial}{\partial x} | g \rangle] / [(E_L - \Delta E_{ig})(E_L - \hbar \omega_{ph} - \Delta E_{i'g})]$  $\hat{e}_i(\updownarrow X)$, $\hat{e}_s(\updownarrow Y)$. This suggests that XY and similarly YX component of Raman tensor changes with change in $E_L$ and this may give rise to change in a different polar dependence of the intensity. Also electron-phonon matrix element $\langle i' | H_{el-ph} | i \rangle$ may also change. When energy of the incident photon changes, then the intermediate state $|i\rangle, |i'\rangle$ changes to new states $|i_{new}\rangle$ and this may also change the denominator. In case of TMDCs this renormalization of the intermediates states may happen easily because of the existence of many energy bands with different symmetries within a short range of energy. Importantly as one moves from bulk to single layer in TMDCs, band topology near the valance band and bottom of conduction band change drastically, and this may also give the different polar plots for a given Raman active phonon modes simply by changing the thickness. We note that similar anisotropic phonon modes as a function of layer thickness and different $E_L$ have been reported in other class of 2D materials with low crystalline symmetry such as black phosphorus, GaTe, ReSe$_2$, ReS$_2$ [29-33].

Our observation of anisotropic polarization dependence for three different forms of MoS$_2$ flakes as a function of thickness and incident photon energy is attributed to the selection rule for optical transition, which can be expressed in quantum theory of Raman scattering via electron-photon-phonon matrix elements as $\langle i|\nabla|g\rangle$, $\langle g|H_{op}|i'\rangle$ and $\langle i'|H_{el-ph}|i\rangle$. Layered TMDCs have large number of conduction band with different symmetries at the $\Gamma$ point in BZ. When incident photon energy or layer thickness is changed then intermediate states gets modified which is reflected in the renormalized electron-photon matrix elements. To get the finite contribution for Raman intensity, both electron-photon matrix elements in eq$^n$. 1 should be finite, which constraint that symmetries of $|i\rangle$, $|i'\rangle$ be linked with $|g\rangle$ state. Based on the group theory one can find the allowed symmetries of the intermediate electronic states $|i\rangle$ and $|i'\rangle$. A detailed list of the intermediate states for a given polarization direction and phonon mode is given in table-S2 and S3 (see suppl. Information). Our observation clearly revealed the strong in-plane anisotropic behavior in high symmetric MoS$_2$. It has been observed that for many low-symmetry 2D systems [29-33] Raman active phonon modes do show polarization dependence on the crystalline orientation. However, it was suggested that polarization dependence as well as anisotropic transport properties such as mobility, is also coupled to the underlying electronic symmetry along with the crystalline symmetry of the material [29-34]. As electron-phonon interaction is an important factor for tuning the transport, thermal and superconductive properties in any material, our findings provide crucial information for the study of anisotropic 2D materials. For example, the out-of-plane vibration ($A_{1g}$) shows much weaker polarization dependence as compared to the in-plane vibration ($E^1_{2g}$), suggesting that the out-of-plane vibrations couple less anisotropically with the in-plane electronic states. Also, the dependence of the modes intensity on laser energy and flake thickness suggest that electrons couple strongly to the $E^1_{2g}$ phonon. We note that anisotropies in transport as well as electronic properties were predicted and reported for the case of MoS$_2$ [35-38]. Our anisotropic results on this high crystalline symmetric material suggest that more theoretical as well experimental studies are required for low crystalline symmetric 2D materials for understanding their underlying properties for the potential future applications.

In summary, we performed a comprehensive Raman scattering studies on three different forms of layered MoS$_2$ probing the anisotropic behavior of the Raman active phonon modes revealing strong anisotropic electron-photon-phonon coupling. The observed anisotropy in the Raman intensity depends strongly on the incident photon energy, phonon energy as well as the thickness of MoS$_2$ for all three different forms studied here. Our work cast a crucial light on understanding the anisotropic strong light-matter interactions in high symmetric MoS$_2$ especially for the in-plane vibrations, and we hope our studies also pave way for the applications of this system in electronic and optoelectronic devices.

**Acknowledgement:** PK acknowledge DST Nano Mission, India, for the financial support. The authors at Mandi acknowledge IIT Mandi for providing the experimental facilities.

**FIGURE CAPTION:**

**FIGURE 1:** (Color online) (a, b) Feynman diagram and schematic representation for the Stokes Raman scattering process, respectively. $\hbar\omega_{in}/\hbar\omega_s$ and $\hbar\omega_{ph}$ corresponds to the energy of incident/scattered photon and phonon energy, respectively. $|i\rangle$ and $|i'\rangle$ represent the intermediate states whereas $|g\rangle$ stands for the ground state. $\langle i|H_{op}|g\rangle$ and $\langle g|H_{op}|i'\rangle$ are the matrix elements for the process of optical absorption and emission, respectively; and $\langle i'|H_{el-ph}|i\rangle$ is the electron-phonon matrix element. (c) Schematic representation of the experimental setup for the Raman measurement. (d) Schematic showing the direction of polarization of the incident ($\hat{e}_i$) and scattered light ($\hat{e}_s$), making angle $\alpha$ and $\alpha+\theta$ with X-axis, respectively.

**FIGURE 2:** (Color online) Raman spectra for four different flakes of horizontally aligned CVD grown $MoS_2$ under two different (a) 532 nm and (b) 633 nm laser excitation energies. The solid red line shows the total sum of Lorentizian fit and thin blue lines show the individual fit of the phonon modes. Flake numbering (from 1 to 4) is done in the increasing order of flakes thickness. Inset show optical image of the region where Raman spectra were measured. $\Delta\omega$ corresponds to the frequency difference between $A_{1g}$ and $E^1_{2g}$ mode.

**FIGURE 3:** (Color online) Intensity polar plots for the Raman active phonon modes $E^1_{2g}$ and $A_{1g}$ with two different laser excitation energies (532 and 633 nm) in horizontally aligned CVD grown

flakes of MoS$_2$. Black spheres are the experimental data points for different polarization angle from $\theta = 0^0$ to $360^0$ and solid blue lines show fitted curves as described in the text. In plane vibrational mode $E_{2g}^1$ shows remarkable anisotropic response by changing the incident photon energy from 532 to 633 nm.

**FIGURE 4:** (Color online) Intensity ratio of $E^1_{2g}$ with respect to $A_{1g}$ in horizontally aligned CVD grown flakes of MoS$_2$ under two different (a) 532nm (b) 633nm laser excitation energies. Solid blue lines are guide to the eye.

**FIGURE 5:** (Color online) Intensity polar plots for the Raman active phonon modes $E^1_{2g}$ and $A_{1g}$ with two different laser excitation energies (532 and 633 nm) in vertically aligned CVD grown flakes of MoS$_2$. Black spheres are experimental data points for different polarization angle from $\theta = 0^0$ to $360^0$ and solid blue lines show fitted curves as described in the text.

**FIGURE 6:** (Color online) Intensity polar plots for the Raman active phonon modes $E^1_{2g}$ and $A_{1g}$ with two different laser excitation energies (532 and 633 nm) in mechanically exfoliated flakes of MoS$_2$. Black spheres are experimental data points for different polarization angle from $\theta = 0^0$ to $360^0$ and solid blue lines show fitted curves as described in the text.

**FIGURE 1:**

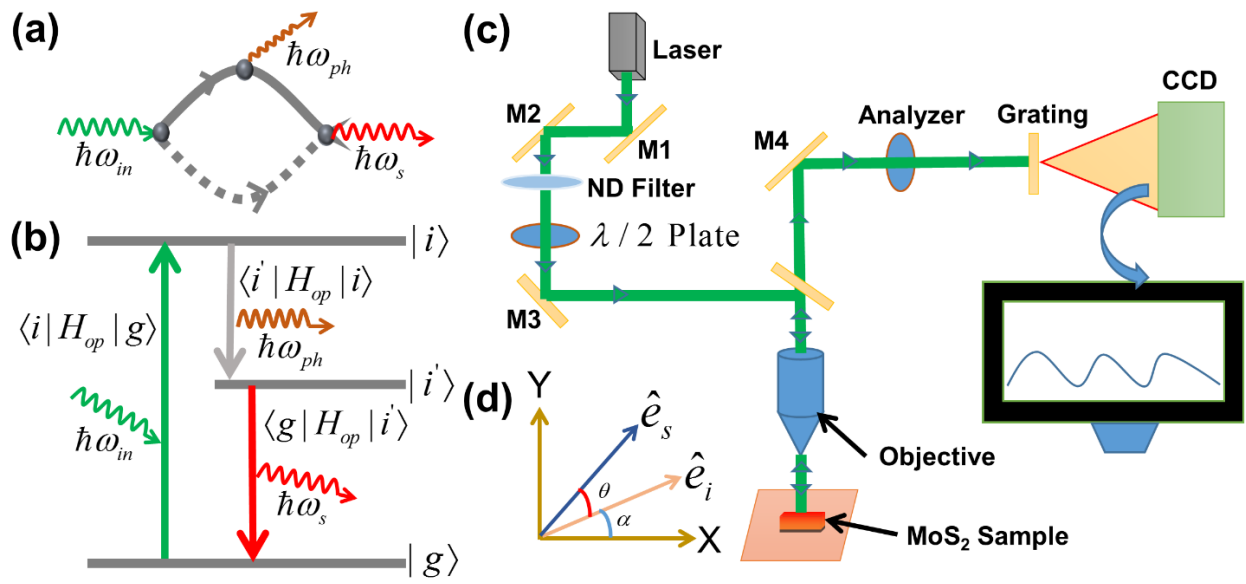

**FIGURE 2:**

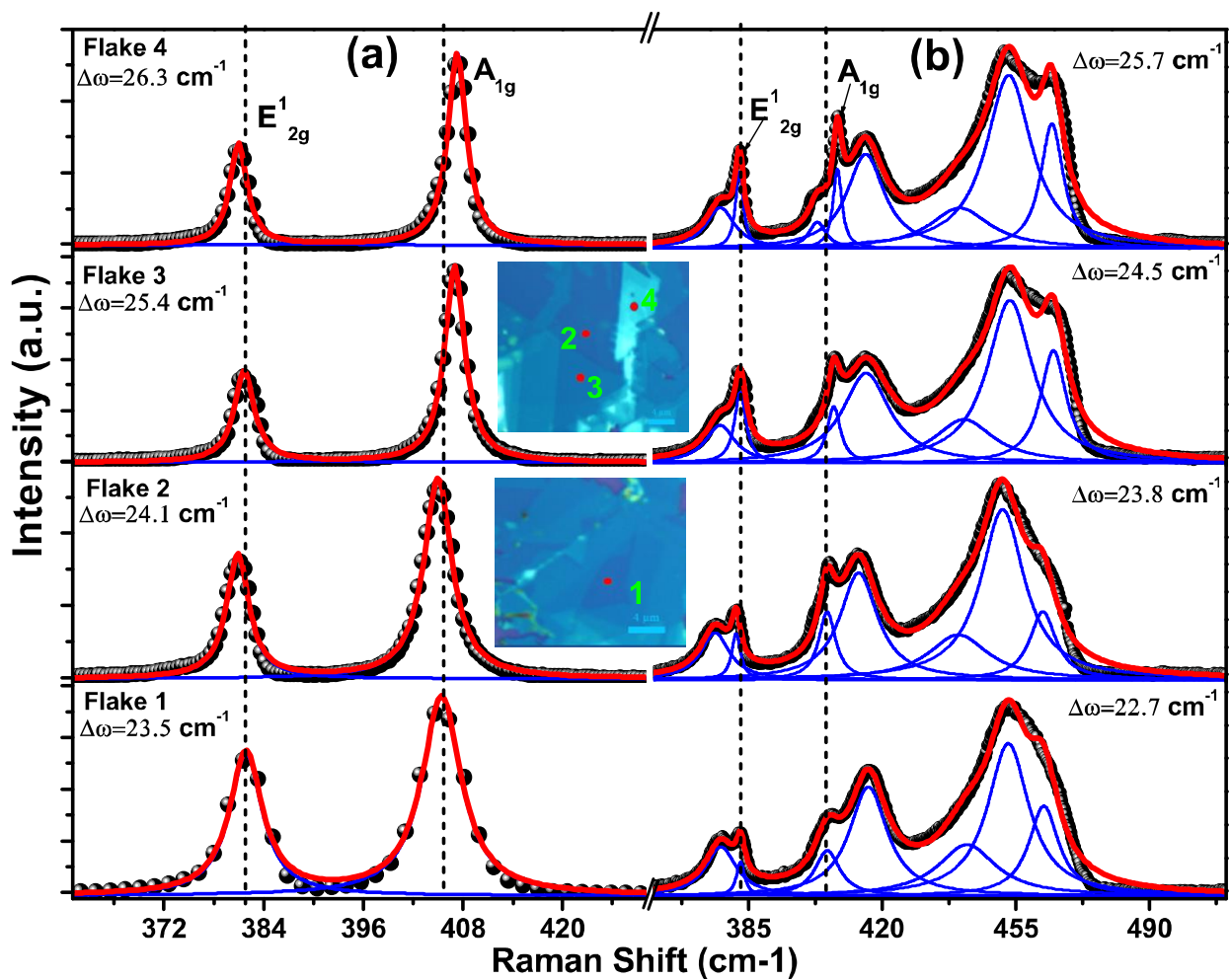

**FIGURE 3:**

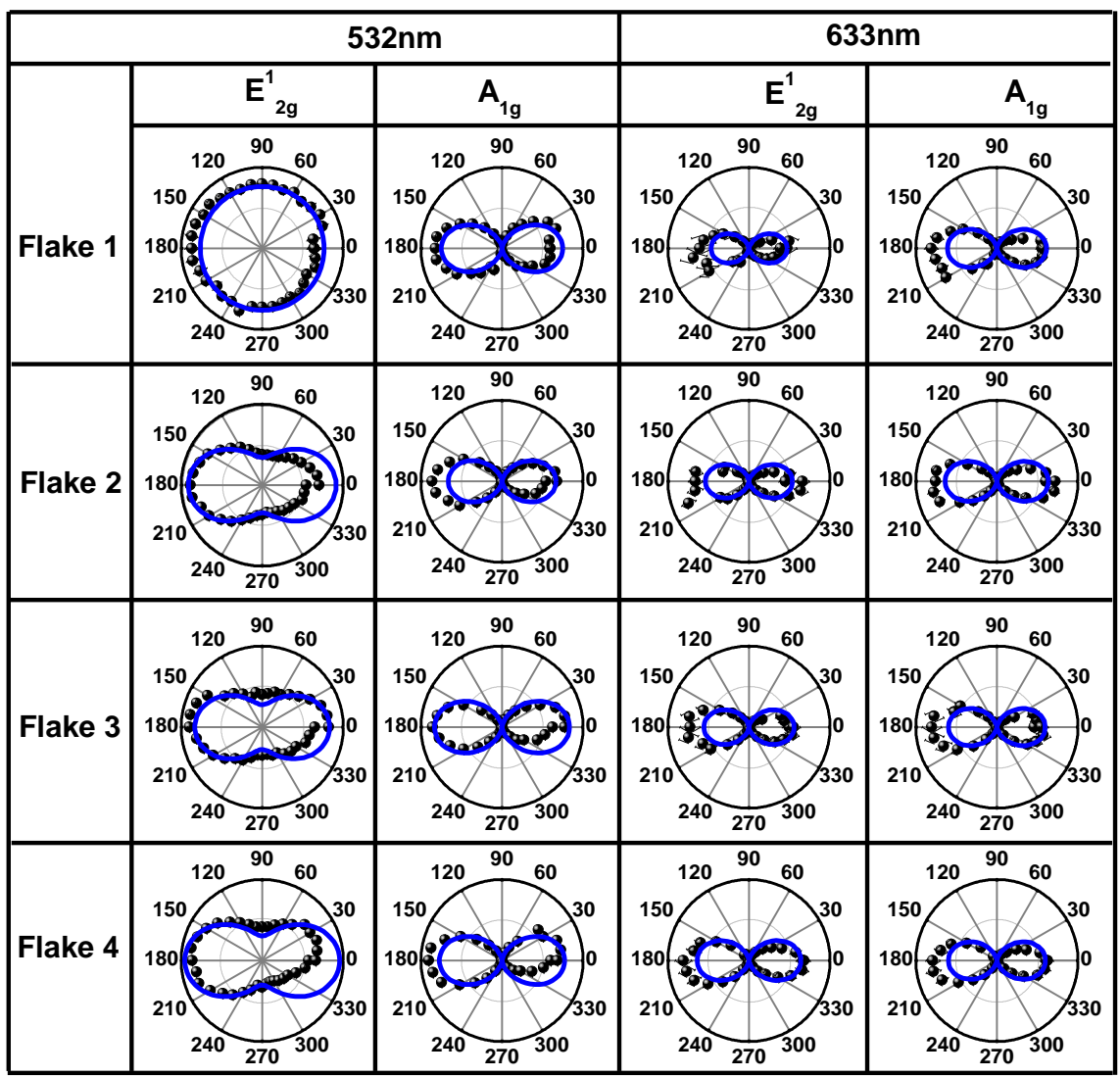

**FIGURE 4:**

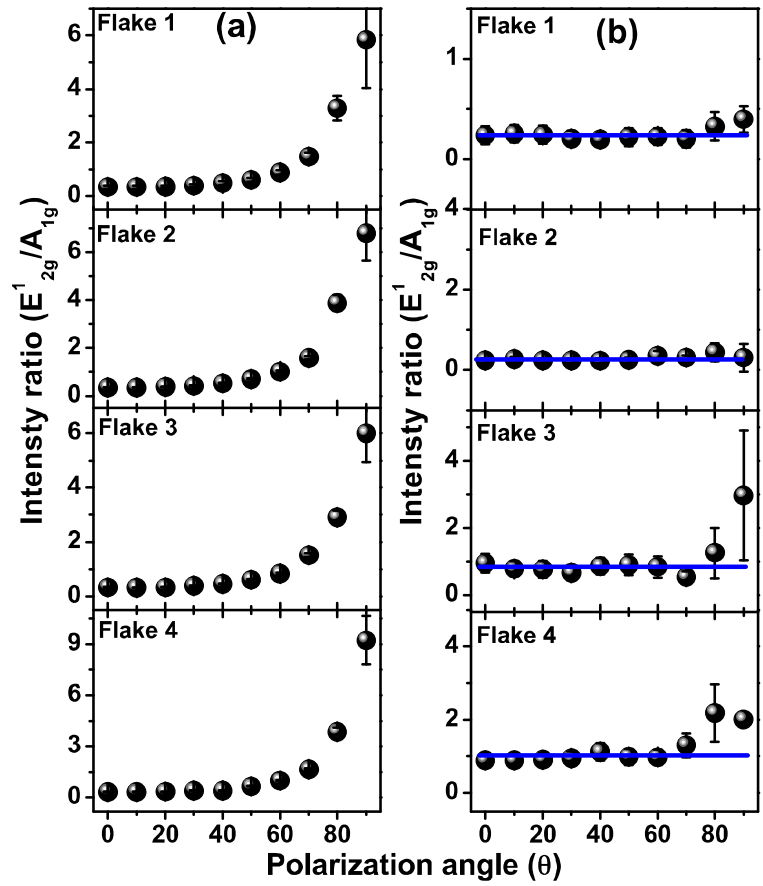

**FIGURE 5:**

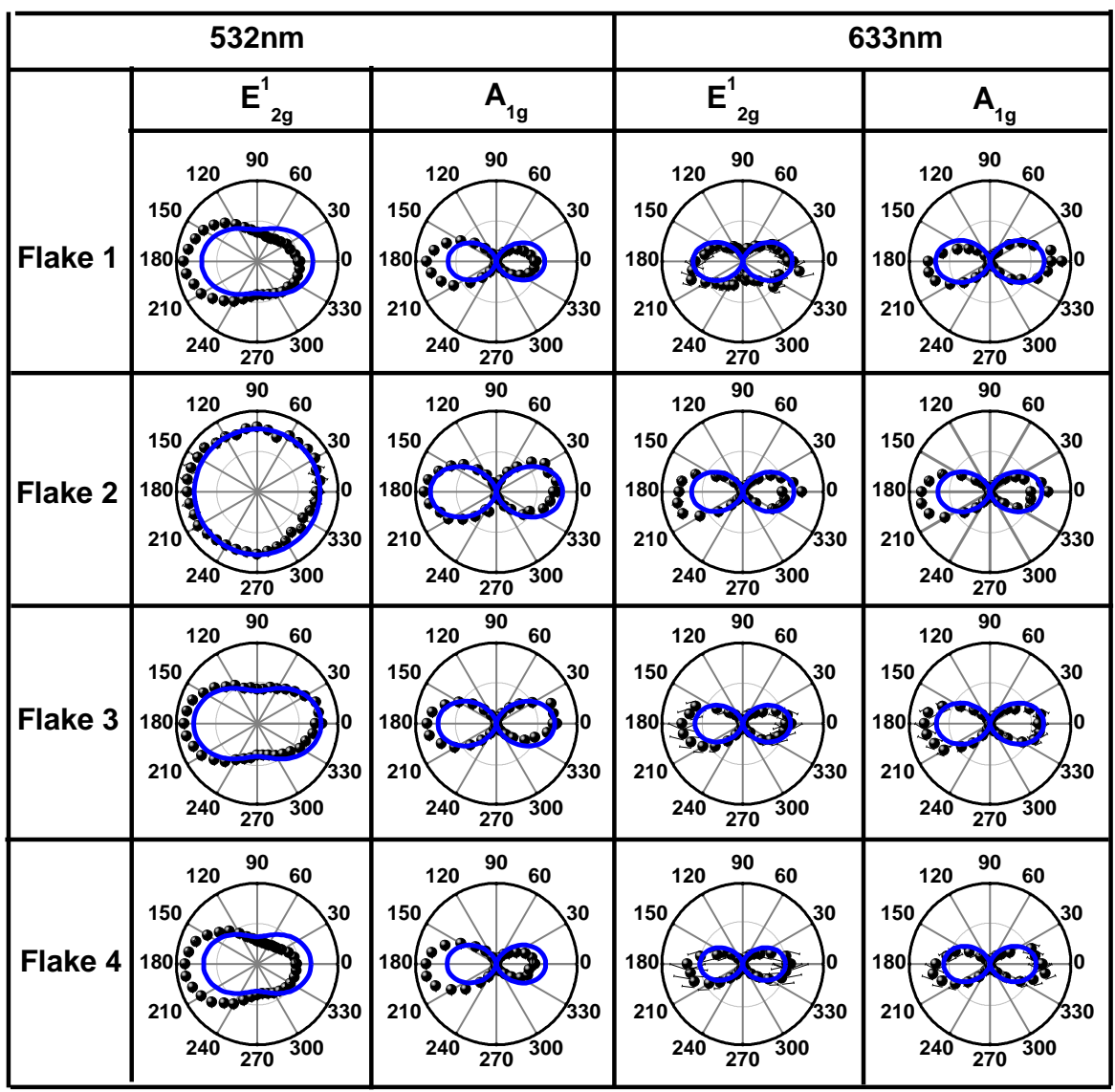

**FIGURE 6:**

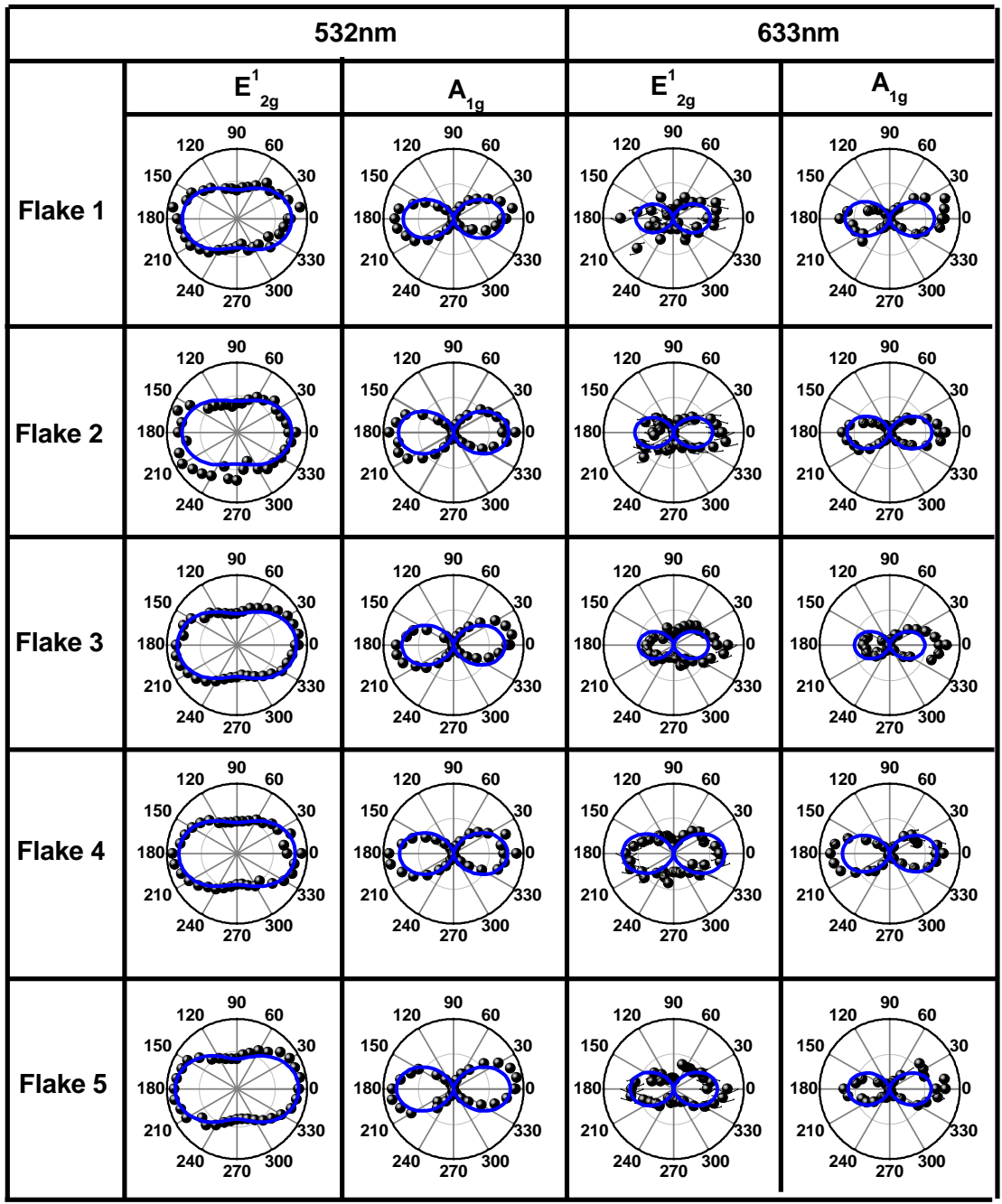

# Supplementary Information:

## Anisotropic Electron-Photon-Phonon Coupling in Layered MoS$_2$

Deepu Kumar[1#], Birender Singh[1], Rahul Kumar[2], Mahesh Kumar[2] and Pradeep Kumar[1*]

[1]School of Basic Sciences, Indian Institute of Technology Mandi, 175005, India

[2]Department of Electrical Engineering, Indian Institute of Technology Jodhpur, Jodhpur-342037, India

#E-mail: deepu7727@gmail.com

* E-mail: pkumar@iitmandi.ac.in

**Space group and phonon modes:**

The bulk 2H-MoS$_2$ belongs to $D_{6h}^4$ point group symmetry ($P6_3mmc$, #194) with two formula units per unit cell (i.e. two MX$_2$ units, 6 atoms). The vibrational degrees of freedom are 18, giving the character table (see table-S1) and irreducible representation of the phonon at the $\Gamma$ as follow:

$\Gamma = A_{1g} + 2A_{2u} + B_{1u} + 2B_{2g} + E_{1g} + 2E_{1u} + E_{2u} + 2E_{2g}$  ($\Gamma_{Raman} = A_{1g} + E_{1g} + 2E_{2g}$, $\Gamma_{IR} = A_{2u} + E_{1u}$, $\Gamma_{Acoustic} = A_{2u} + E_{1u}$, $\Gamma_{Silent} = B_{1u} + 2B_{2g} + E_{2u}$) [1-3]. There are two kinds of $E_{2g}$ modes labelled as $E_{2g}^1$ and $E_{2g}^2$; $E_{2g}^1$ is associated with the in-plane vibrations of Mo and S atoms in opposite directions and $E_{2g}^2$ is a low frequency interlayer mode hence is not observed in case of monolayer. Monolayer/odd number layers of MoS$_2$ belongs to point group $D_{3h}^1$ and space group ($P\bar{6}m2$, #187) symmetry and there are three number of atoms per unit cell i.e. one metal (Mo) atom and 2 sulfur (S) atoms giving 9 normal mode of vibrations at $\Gamma$ point of BZ with the following irreducible representation $\Gamma = A_1' + 2E' + 2A_2'' + E''$ ($\Gamma_{Raman} = A_1' + E' + E''$, $\Gamma_{IR} = E' + A_2''$, $\Gamma_{Acoustic} = E' + A_2''$). However, bilayer/even number layers of MoS$_2$ belongs to $D_{3d}^3(P\bar{3}m1, \#164)$ symmetry and there are 6 number of atoms per unit cell giving 18 normal mode of vibration at the $\Gamma$ point and can be expressed into the following irreducible representation $\Gamma = 3A_{1g} + 3E_g + 3A_{2u} + 3E_u$ ($\Gamma_{Raman} = 3A_{1g} + 3E_g$, $\Gamma_{IR} = 2A_{2u} + 2E_u$, $\Gamma_{Silent} = A_{2u} + E_u$). An important difference between these two groups is the presence of inversion symmetry and absence of $\sigma_h$ plane in bilayer/even number of layer system. We note that nomenclature of the phonon modes changes depending upon the odd/even number of layer as $A_{1g} \to A_1'$, $E_{2g} \to E'/E_g$ and $E_{1g} \to E''/E_g$. For convenience, for the Raman active phonon modes, we have used the nomenclature as used for bulk sample i.e. $E^1_{2g}$ and $A_{1g}$, we note that same nomenclature is also used in the literature.

**Table-S1:** Character table for **(a)** $D_{6h}^4$ **(b)** $D_{3h}^1$ and **(c)** $D_{3d}^3$ point groups. There are 24 symmetry operations for $D_{6h}$ and 12 for $D_{3h}^1$ and $D_{3d}^3$. Raman tensor for the Raman active phonon modes for each point group are given at the bottom of each table.

| (a) | E | 2C$_6$ | 2C$_3$ | C$_2$ | 3C'$_2$ | 3C''$_2$ | i | 2S$_3$ | 2S$_6$ | $\sigma_h$ | 3$\sigma_d$ | 3$\sigma_v$ | Linear, rotations | Quadratic |
|---|---|---|---|---|---|---|---|---|---|---|---|---|---|---|
| $A_{1g}$ | 1 | 1 | 1 | 1 | 1 | 1 | 1 | 1 | 1 | 1 | 1 | 1 | | $x^2+y^2$, $z^2$ |
| $A_{2g}$ | 1 | 1 | 1 | 1 | -1 | -1 | 1 | 1 | 1 | 1 | -1 | -1 | R$_z$ | |
| $B_{1g}$ | 1 | -1 | 1 | -1 | 1 | -1 | 1 | -1 | 1 | -1 | 1 | -1 | | |
| $B_{2g}$ | 1 | -1 | 1 | -1 | -1 | 1 | 1 | -1 | 1 | -1 | -1 | 1 | | |
| $E_{1g}$ | 2 | 1 | -1 | -2 | 0 | 0 | 2 | 1 | -1 | -2 | 0 | 0 | (R$_x$, R$_y$) | (xz, yz) |
| $E_{2g}$ | 2 | -1 | -1 | 2 | 0 | 0 | 2 | -1 | -1 | 2 | 0 | 0 | | ($x^2$-$y^2$, xy) |
| $A_{1u}$ | 1 | 1 | 1 | 1 | 1 | 1 | -1 | -1 | -1 | -1 | -1 | -1 | | |
| $A_{2u}$ | 1 | 1 | 1 | 1 | -1 | -1 | -1 | -1 | -1 | -1 | 1 | 1 | z | |
| $B_{1u}$ | 1 | -1 | 1 | -1 | 1 | -1 | -1 | 1 | -1 | 1 | -1 | 1 | | |
| $B_{2u}$ | 1 | -1 | 1 | -1 | -1 | 1 | -1 | 1 | -1 | 1 | 1 | -1 | | |

| | E | $2C_6$ | $2C_3$ | $C_2$ | $3C_2'$ | $3C_2''$ | i | $2S_3$ | $2S_6$ | $\sigma_h$ | $3\sigma_d$ | $3\sigma_v$ | Linear, rotations | Quadratic |
|---|---|---|---|---|---|---|---|---|---|---|---|---|---|---|
| $E_{1u}$ | 2 | 1 | -1 | -2 | 0 | 0 | -2 | -1 | 1 | 2 | 0 | 0 | (x, y) | |
| $E_{2u}$ | 2 | -1 | -1 | 2 | 0 | 0 | -2 | 1 | 1 | -2 | 0 | 0 | | |

$\Gamma_{Raman} = A_{1g} + E_{1g} + 2E_{2g}$

$$A_{1g} = \begin{pmatrix} a & 0 & 0 \\ 0 & a & 0 \\ 0 & 0 & b \end{pmatrix} \quad E_{1g} = \begin{pmatrix} 0 & 0 & 0 \\ 0 & 0 & c \\ 0 & c & 0 \end{pmatrix}, \begin{pmatrix} 0 & 0 & -c \\ 0 & 0 & 0 \\ -c & 0 & 0 \end{pmatrix} E_{2g} = \begin{pmatrix} 0 & d & 0 \\ d & 0 & 0 \\ 0 & 0 & 0 \end{pmatrix}, \begin{pmatrix} d & 0 & 0 \\ 0 & -d & 0 \\ 0 & 0 & 0 \end{pmatrix}$$

| (b) | E | $2C_3$ | $3C_2$ | $\sigma_h$ | $2S_3$ | $3\sigma_v$ | Linear, rotations | Quadratic |
|---|---|---|---|---|---|---|---|---|
| $A_1'$ | 1 | 1 | 1 | 1 | 1 | 1 | | $x^2+y^2, z^2$ |
| $A_2'$ | 1 | 1 | -1 | 1 | 1 | -1 | $R_z$ | |
| $E'$ | 2 | -1 | 0 | 2 | -1 | 0 | (x, y) | $(x^2-y^2, xy)$ |
| $A_1''$ | 1 | 1 | 1 | -1 | -1 | -1 | | |
| $A_2''$ | 1 | 1 | -1 | -1 | -1 | 1 | z | |
| $E''$ | 2 | -1 | 0 | -2 | 1 | 0 | $(R_x, R_y)$ | (xz, yz) |

$\Gamma_{Raman} = A_1' + E' + E''$

$$A_1' = \begin{pmatrix} a & 0 & 0 \\ 0 & a & 0 \\ 0 & 0 & b \end{pmatrix} \quad E' = \begin{pmatrix} 0 & d & 0 \\ d & 0 & 0 \\ 0 & 0 & 0 \end{pmatrix}, \begin{pmatrix} d & 0 & 0 \\ 0 & -d & 0 \\ 0 & 0 & 0 \end{pmatrix} \quad E'' = \begin{pmatrix} 0 & 0 & 0 \\ 0 & 0 & c \\ 0 & c & 0 \end{pmatrix}, \begin{pmatrix} 0 & 0 & -c \\ 0 & 0 & 0 \\ -c & 0 & 0 \end{pmatrix}$$

| (c) | E | $2C_3$ | $3C_2$ | i | $2S_6$ | $3\sigma_d$ | Linear, rotations | Quadratic |
|---|---|---|---|---|---|---|---|---|
| $A_{1g}$ | 1 | 1 | 1 | 1 | 1 | 1 | | $x^2+y^2, z^2$ |
| $A_{2g}$ | 1 | 1 | -1 | 1 | 1 | -1 | $R_z$ | |
| $E_g$ | 2 | -1 | 0 | 2 | -1 | 0 | $(R_x, R_y)$ | $(x^2-y^2, xy)$ (xz, yz) |
| $A_{1u}$ | 1 | 1 | 1 | -1 | -1 | -1 | | |
| $A_{2u}$ | 1 | 1 | -1 | -1 | -1 | 1 | z | |
| $E_u$ | 2 | -1 | 0 | -2 | 1 | 0 | (x, y) | |

$\Gamma_{Raman} = 3A_{1g} + 3E_g$

$$A_{1g} = \begin{pmatrix} a & 0 & 0 \\ 0 & a & 0 \\ 0 & 0 & b \end{pmatrix} \quad E_g = \begin{pmatrix} c & 0 & 0 \\ 0 & -c & d \\ 0 & d & 0 \end{pmatrix}, \begin{pmatrix} 0 & -c & -d \\ -c & 0 & 0 \\ -d & 0 & 0 \end{pmatrix}$$

**Selection rules for optical transitions:**

Polarization dependence of the intensity for a given Raman active phonon mode may be understood by using constraint owing to the group theory selection rule for selecting the intermediate states $|i\rangle, |i'\rangle$. From eq$^n$. 1 (given in main text), condition for obtaining the finite intensity for a given phonon mode, say $A_{1g}$ phonon mode, is that the total product of the symmetry operation should be such that $\Gamma_{op}^{fi'} \otimes \Gamma_{el-ph}^{i'i} \otimes \Gamma_{op}^{ig} \subset A_{1g}$, where $\Gamma_{op}^{fi'}, \Gamma_{op}^{ig}$ and $\Gamma_{el-ph}^{i'i}$ are the irreducible representations of the dipole vector component parallel to $\hat{e}_s$, $\hat{e}_i$ and electron-phonon interaction, respectively. Table-S2 and S3 gives the details of the intermediate states involved in the optical transitions for all the three groups for bulk, odd and even number of layer; concerning for the observed Raman active phonon modes (i.e. $A_{1g}$ and $E_{2g}^1$).

**Table-S2:** Selection rules for the intermediate states $|i\rangle, |i'\rangle$ for a given ground state $|g\rangle|$ for point group $D_{6h}^4$, **(a)** for polarization vector xx or yy (i.e. $\hat{e}_i$ is parallel to $\hat{e}_s$) which both correspond to the $A_{1g}$ phonon excitation and **(b)** $E_{2g}$ phonon for xx/yy and xy polarization (i.e. $\hat{e}_i$ is either parallel or perpendicular to $\hat{e}_s$). These selection rules correspond to the following product of matrix elements: $\langle f|H_{op}|i'\rangle\langle i'|H_{el-ph}|i\rangle\langle i|H_{op}|g\rangle$ with $\langle f|=|g\rangle$

| (a) | XX | | | YY | | |
|---|---|---|---|---|---|---|
| $|g\rangle$ | $|i\rangle$ | $|i'\rangle$ | $|g\rangle$ | $|i\rangle$ | $|i'\rangle$ |
| $A_{1g}$ | $E_{1u}$ | $E_{1u}$ | $A_{1g}$ | $E_{1u}$ | $E_{1u}$ |
| $A_{2g}$ | $E_{1u}$ | $E_{1u}$ | $A_{2g}$ | $E_{1u}$ | $E_{1u}$ |
| $B_{1g}$ | $E_{2u}$ | $E_{2u}$ | $B_{1g}$ | $E_{2u}$ | $E_{2u}$ |
| $B_{2g}$ | $E_{2u}$ | $E_{2u}$ | $B_{2g}$ | $E_{2u}$ | $E_{2u}$ |
| $E_{1g}$ | $A_{1u}+A_{2u}+E_{2u}$ | $A_{1u}+A_{2u}+E_{2u}$ | $E_{1g}$ | $A_{1u}+A_{2u}+E_{2u}$ | $A_{1u}+A_{2u}+E_{2u}$ |
| $E_{2g}$ | $B_{1u}+B_{2u}+E_{1u}$ | $B_{1u}+B_{2u}+E_{1u}$ | $E_{2g}$ | $B_{1u}+B_{2u}+E_{1u}$ | $B_{1u}+B_{2u}+E_{1u}$ |
| $A_{1u}$ | $E_{1g}$ | $E_{1g}$ | $A_{1u}$ | $E_{1g}$ | $E_{1g}$ |
| $A_{2u}$ | $E_{1g}$ | $E_{1g}$ | $A_{2u}$ | $E_{1g}$ | $E_{1g}$ |
| $B_{1u}$ | $E_{2g}$ | $E_{2g}$ | $B_{1u}$ | $E_{2g}$ | $E_{2g}$ |
| $B_{2u}$ | $E_{2g}$ | $E_{2g}$ | $B_{2u}$ | $E_{2g}$ | $E_{2g}$ |
| $E_{1u}$ | $A_{1g}+A_{2g}+E_{2g}$ | $A_{1g}+A_{2g}+E_{2g}$ | $E_{1u}$ | $A_{1g}+A_{2g}+E_{2g}$ | $A_{1g}+A_{2g}+E_{2g}$ |
| $E_{2u}$ | $B_{1g}+B_{2g}+E_{1g}$ | $B_{1g}+B_{2g}+E_{1g}$ | $E_{2u}$ | $B_{1g}+B_{2g}+E_{1g}$ | $B_{1g}+B_{2g}+E_{1g}$ |

| (b) | XX/YY | | | XY | | |
|---|---|---|---|---|---|---|
| $|g\rangle$ | $|i\rangle$ | $|i'\rangle$ | $|g\rangle$ | $|i\rangle$ | $|i'\rangle$ | |
| $A_{1g}$ | $E_{1u}$ | $E_{1u}$ | $A_{1g}$ | $E_{1u}$ | $E_{1u}$ | |
| $A_{2g}$ | $E_{1u}$ | $E_{1u}$ | $A_{2g}$ | $E_{1u}$ | $E_{1u}$ | |
| $B_{1g}$ | $E_{2u}$ | $E_{2u}$ | $B_{1g}$ | $E_{2u}$ | $E_{2u}$ | |
| $B_{2g}$ | $E_{2u}$ | $E_{2u}$ | $B_{2g}$ | $E_{2u}$ | $E_{2u}$ | |
| $E_{1g}$ | $A_{1u}+A_{2u}+E_{2u}$ | $A_{1u}+A_{2u}+E_{2u}$ | $E_{1g}$ | $A_{1u}+A_{2u}+E_{2u}$ | $A_{1u}+A_{2u}+E_{2u}$ | |
| $E_{2g}$ | $B_{1u}+B_{2u}+E_{1u}$ | $B_{1u}+B_{2u}+E_{1u}$ | $E_{2g}$ | $B_{1u}+B_{2u}+E_{1u}$ | $B_{1u}+B_{2u}+E_{1u}$ | |
| $A_{1u}$ | $E_{1g}$ | $E_{1g}$ | $A_{1u}$ | $E_{1g}$ | $E_{1g}$ | |
| $A_{2u}$ | $E_{1g}$ | $E_{1g}$ | $A_{2u}$ | $E_{1g}$ | $E_{1g}$ | |
| $B_{1u}$ | $E_{2g}$ | $E_{2g}$ | $B_{1u}$ | $E_{2g}$ | $E_{2g}$ | |
| $B_{2u}$ | $E_{2g}$ | $E_{2g}$ | $B_{2u}$ | $E_{2g}$ | $E_{2g}$ | |
| $E_{1u}$ | $A_{1g}+A_{2g}+E_{2g}$ | $A_{1g}+A_{2g}+E_{2g}$ | $E_{1u}$ | $A_{1g}+A_{2g}+E_{2g}$ | $A_{1g}+A_{2g}+E_{2g}$ | |
| $E_{2u}$ | $B_{1g}+B_{2g}+E_{1g}$ | $B_{1g}+B_{2g}+E_{1g}$ | $E_{2u}$ | $B_{1g}+B_{2g}+E_{1g}$ | $B_{1g}+B_{2g}+E_{1g}$ | |

**Table-S3:** Selection rules for the intermediate states $|i\rangle, |i'\rangle$ for a given ground state $|g\rangle|$ for point group $D_{3h}^1$ and $D_{3D}^3$ (a) for polarization vector xx or yy which both correspond to the $A_1'$ phonon excitation, (b) $E'$ phonon for xx/yy and xy polarization, (c) for polarization vector xx or yy which both correspond to the $A_{1g}$ phonon excitation, and (d) $E_g$ phonon for xx/yy and xy polarization. These selection rules correspond to the following product of matrix elements: $\langle f|H_{op}|i'\rangle\langle i'|H_{el-ph}|i\rangle\langle i|H_{op}|g\rangle$ with $\langle f|=|g\rangle$

| (a) | XX | | | YY | | |
|---|---|---|---|---|---|---|
| $|g\rangle$ | $|i\rangle$ | $|i'\rangle$ | $|g\rangle$ | $|i\rangle$ | $|i'\rangle$ |
| $A_1'$ | $E'$ | $E'$ | $A_1'$ | $E'$ | $E'$ |
| $A_2'$ | $E'$ | $E'$ | $A_2'$ | $E'$ | $E'$ |
| $E'$ | $A_1'+A_2'+E'$ | $A_1'+A_2'+E'$ | $E'$ | $A_1'+A_2'+E'$ | $A_1'+A_2'+E'$ |
| $A_1''$ | $E''$ | $E''$ | $A_1''$ | $E''$ | $E''$ |
| $A_2''$ | $E''$ | $E''$ | $A_2''$ | $E''$ | $E''$ |
| $E''$ | $A_1''+A_1''+E''$ | $A_1''+A_1''+E''$ | $E''$ | $A_1''+A_1''+E''$ | $A_1''+A_1''+E''$ |

| (b) | XX/YY | | | XY | | |
|---|---|---|---|---|---|---|
| $|g\rangle$ | $|i\rangle$ | $|i'\rangle$ | $|g\rangle$ | $|i\rangle$ | $|i'\rangle$ |
| $A_1'$ | $E'$ | $E'$ | $A_1'$ | $E'$ | $E'$ |
| $A_2'$ | $E'$ | $E'$ | $A_2'$ | $E'$ | $E'$ |
| $E'$ | $A_1'+A_2'+E'$ | $A_1'+A_2'+E'$ | $E'$ | $A_1'+A_2'+E'$ | $A_1'+A_2'+E'$ |
| $A_1''$ | $E''$ | $E''$ | $A_1''$ | $E''$ | $E''$ |
| $A_2''$ | $E''$ | $E''$ | $A_2''$ | $E''$ | $E''$ |
| $E''$ | $A_1''+A_1''+E''$ | $A_1''+A_1''+E''$ | $E''$ | $A_1''+A_1''+E''$ | $A_1''+A_1''+E''$ |

| (c) | XX | | | YY | | |
|---|---|---|---|---|---|---|
| $|g\rangle$ | $|i\rangle$ | $|i'\rangle$ | $|g\rangle$ | $|i\rangle$ | $|i'\rangle$ | |
| $A_{1g}$ | $E_u$ | $E_u$ | $A_{1g}$ | $E_u$ | $E_u$ | |
| $A_{2g}$ | $E_u$ | $E_u$ | $A_{2g}$ | $E_u$ | $E_u$ | |
| $E_g$ | $A_{1u} + A_{2u} + E_u$ | $A_{1u} + A_{2u} + E_u$ | $E_g$ | $A_{1u} + A_{2u} + E_u$ | $A_{1u} + A_{2u} + E_u$ | |
| $A_{1u}$ | $E_g$ | $E_g$ | $A_{1u}$ | $E_g$ | $E_g$ | |
| $A_{2u}$ | $E_g$ | $E_g$ | $A_{2u}$ | $E_g$ | $E_g$ | |
| $E_u$ | $A_{1g} + A_{2g} + E_g$ | $A_{1g} + A_{2g} + E_g$ | $E_u$ | $A_{1g} + A_{2g} + E_g$ | $A_{1g} + A_{2g} + E_g$ | |

| (d) | XX/XY | | | XY | | |
|---|---|---|---|---|---|---|
| $|g\rangle$ | $|i\rangle$ | $|i'\rangle$ | $|g\rangle$ | $|i\rangle$ | $|i'\rangle$ | |
| $A_{1g}$ | $E_u$ | $E_u$ | $A_{1g}$ | $E_u$ | $E_u$ | |
| $A_{2g}$ | $E_u$ | $E_u$ | $A_{2g}$ | $E_u$ | $E_u$ | |
| $E_g$ | $A_{1u} + A_{2u} + E_u$ | $A_{1u} + A_{2u} + E_u$ | $E_g$ | $A_{1u} + A_{2u} + E_u$ | $A_{1u} + A_{2u} + E_u$ | |
| $A_{1u}$ | $E_g$ | $E_g$ | $A_{1u}$ | $E_g$ | $E_g$ | |
| $A_{2u}$ | $E_g$ | $E_g$ | $A_{2u}$ | $E_g$ | $E_g$ | |
| $E_u$ | $A_{1g} + A_{2g} + E_g$ | $A_{1g} + A_{2g} + E_g$ | $E_u$ | $A_{1g} + A_{2g} + E_g$ | $A_{1g} + A_{2g} + E_g$ | |

**Figure S1**: (Color online) Raman spectra for four different flakes in vertically aligned CVD grown $MoS_2$ under two different (a) 532nm and (b) 633nm laser excitation energies. The solid red line shows the total sum of Lorentizian fit and thin blue lines show individual fit of the phonon modes. Flake numbering (from 1 to 4) is done in the increasing order of flake thickness. Inset show optical image of the region where Raman spectra were measured. $\Delta\omega$ corresponds to the frequency difference between $A_{1g}$ and $E^1_{2g}$ mode.

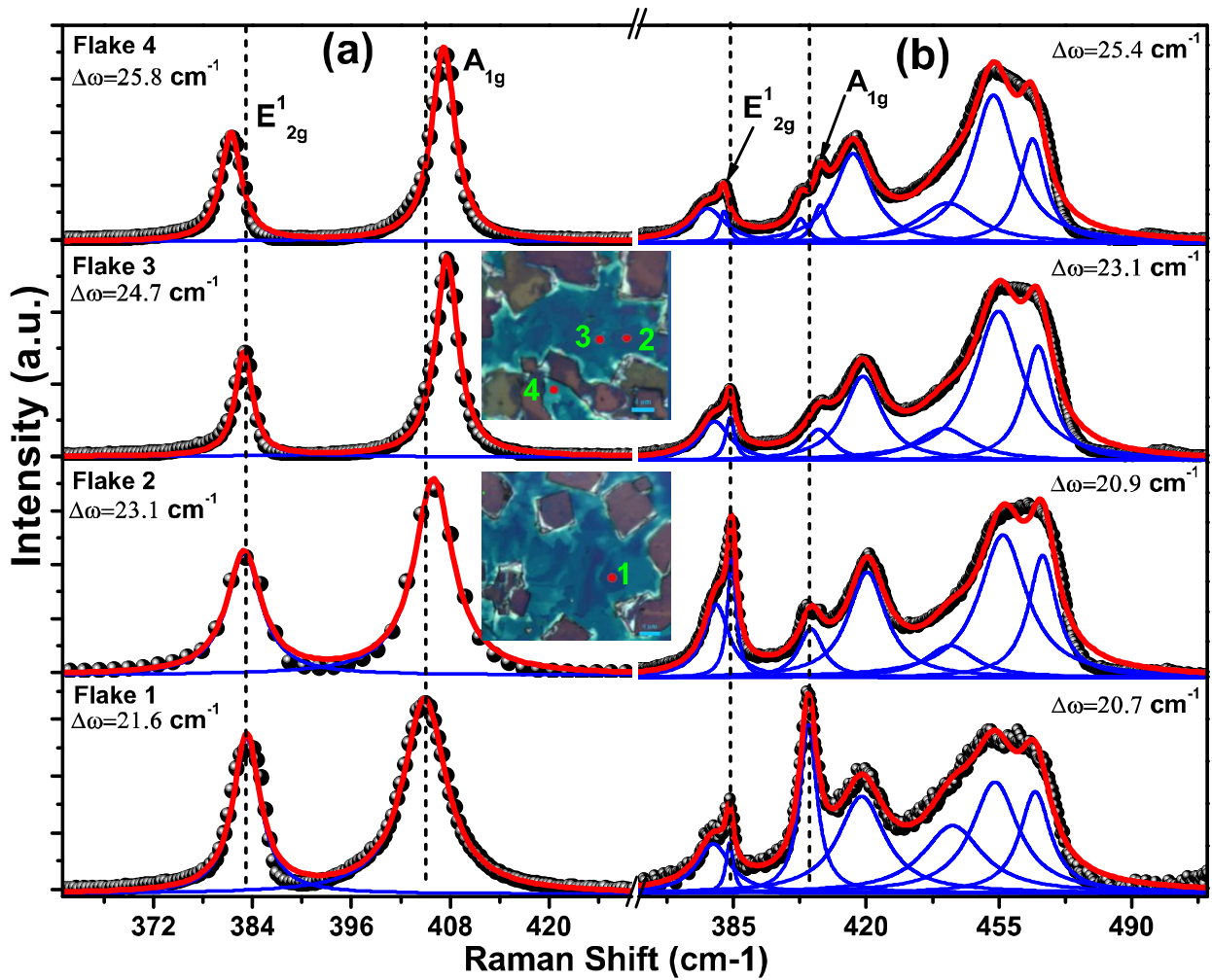

**Figure S2**: (Color online) Intensity ratio of $E^1_{2g}$ mode with respect to $A_{1g}$ in vertically aligned CVD grown $MoS_2$ under two different (a) 532nm (b) 633nm laser excitation energies. Solid blue lines are guide to the eye.

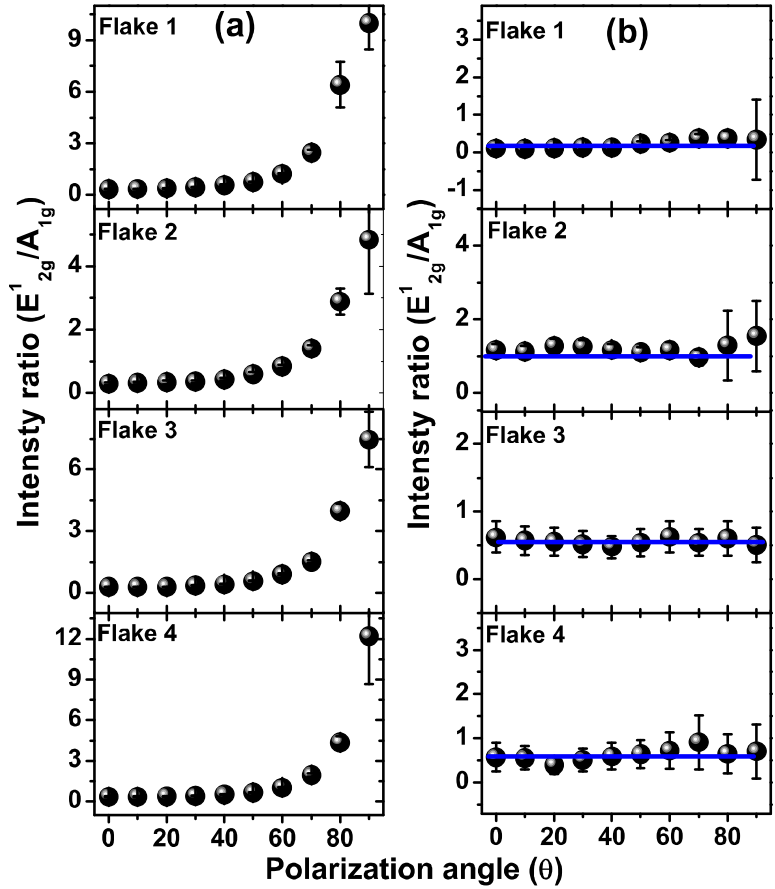

**Figure S3**: Color online) Raman spectra for five different flakes in mechanically exfoliated $MoS_2$ under two different (a) 532nm and (b) 633nm laser excitation energies. The solid red line shows the total sum of Lorentizian fit and thin blue lines show individual fit of the phonon modes. Flake numbering (from 1 to 5) is done in the increasing order of flake thickness. Inset show optical image of the region where Raman spectra were measured. $\Delta\omega$ corresponds to the frequency difference between $A_{1g}$ and $E_{2g}^1$ mode.

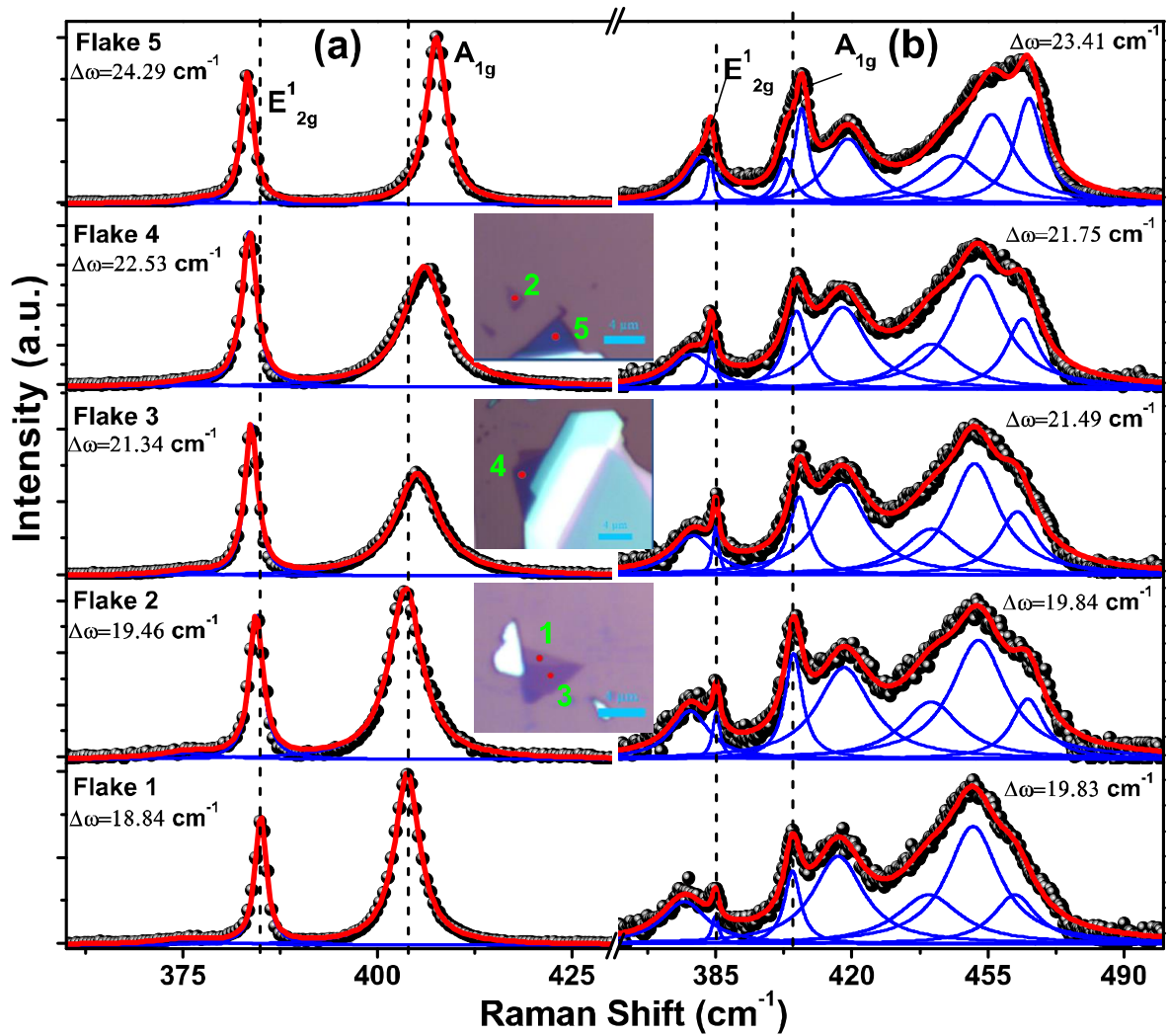

**Figure S4**: (Color online) Intensity ratio of $E^1_{2g}$ mode with respect to $A_{1g}$ in mechanically exfoliated $MoS_2$ under two different (a) 532nm (b) 633nm laser excitation energies. Solid blue lines are guide to the eye.

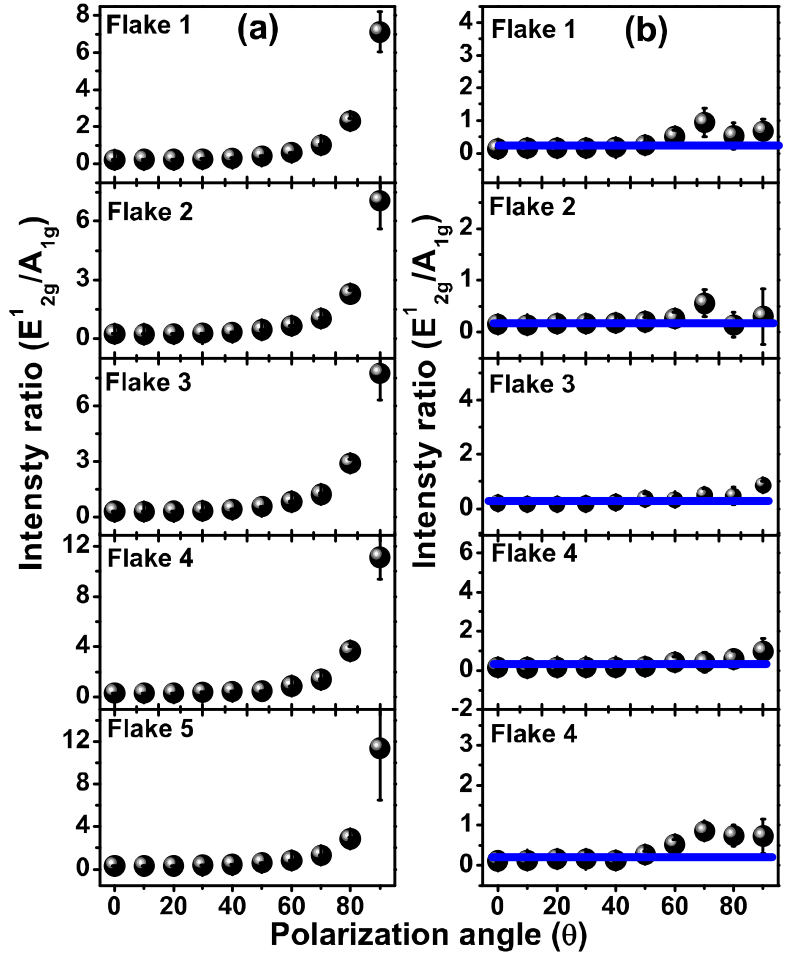